\documentclass[twocolumn,showpacs,aps,pre]{revtex4-1}

\usepackage{amssymb,amsmath}
\usepackage[dvips]{graphicx}

\begin{document}

\title{Rare regions and Griffiths singularities at a clean critical point:
The five-dimensional disordered contact process}

\author{Thomas Vojta}
\affiliation{Department of Physics, Missouri University of Science and Technology,
Rolla, MO 65409, USA}

\author{John Igo}
\affiliation{Department of Physics and Astronomy, Washington State University,
Pullman, WA 99164, USA}
\affiliation{Department of Physics, Missouri University of Science and Technology,
Rolla, MO 65409, USA}

\author{Jos\'e A. Hoyos}
\affiliation{Instituto de F\'{i}sica de S\~ao Carlos, Universidade de S\~ao Paulo,
C.P. 369, S\~ao Carlos, S\~ao Paulo 13560-970, Brazil}

\begin{abstract}
We investigate the nonequilibrium phase transition of the disordered contact process in five
space dimensions by means of optimal fluctuation theory and Monte Carlo simulations.
We find that the critical behavior is of mean-field type, i.e., identical to that of the
clean five-dimensional contact process. It is accompanied by off-critical power-law Griffiths
singularities whose dynamical exponent $z'$ saturates at a finite value as the transition is
approached.
These findings resolve the apparent contradiction between the Harris criterion which implies that weak disorder is
renormalization-group irrelevant and the rare-region classification which predicts unconventional behavior.
 We confirm and illustrate our theory by large-scale Monte-Carlo simulations
of systems with up to $70^5$ sites. We also relate our results to a recently established general
relation between the Harris criterion and Griffiths singularities [Phys. Rev. Lett. {\bf 112}, 075702 (2014)],
and we discuss implications for other phase transitions.
\end{abstract}

\date{\today}
\pacs{05.70.Ln, 64.60.Ht, 02.50.Ey}

\maketitle

%%%%%%%%%%%%%%%%%%%%%%%%%%%%%%%%%%%%%%%%%%%%%%%%%%%%%%%%%%%%%%%%%%%%%%%%%%%%%%%%%
% Main text starts here
%%%%%%%%%%%%%%%%%%%%%%%%%%%%%%%%%%%%%%%%%%%%%%%%%%%%%%%%%%%%%%%%%%%%%%%%%%%%%%%%%

%%%%%%%%%%%%%%%%%%%%%%%%%%%%%%%%%%%%%%%%%%%%%%%%%%%%%%%%%%%%%%%%%%%%%%%%%%%%%%%%%
\section{Introduction}
\label{sec:Intro}
%%%%%%%%%%%%%%%%%%%%%%%%%%%%%%%%%%%%%%%%%%%%%%%%%%%%%%%%%%%%%%%%%%%%%%%%%%%%%%%%%

Over the last several decades, enormous progress has been made in understanding the influence of
quenched random disorder on critical points. Early work focused on thermal (classical) phase transitions
and often used perturbative methods borrowed from the analysis of phase transitions in clean systems
(for an early review, see, e.g., Ref.\ \cite{Grinstein85}). Later work studied disorder effects
at zero-temperature quantum phase transitions as well as nonequilibrium phase transitions. At many of
these transitions, disorder has stronger, non-perturbative effects related to rare, atypically strong
disorder fluctuations (for reviews, see, e.g., Refs.\ \cite{Vojta06,Vojta10}).

From this work, two different frameworks for classifying the effects of disorder on critical points
have emerged. The first classification is based on the behavior of the average disorder strength under
coarse graining \cite{MMHF00}. If (weak) disorder decreases without limit under coarse graining, it becomes
unimportant on the large length scales that govern a critical point. The critical behavior of the
disordered system is therefore identical to the corresponding clean one. According to the Harris criterion
\cite{Harris74}, this case is realized if the correlation length exponent $\nu_\perp$ of the clean system
fulfills the inequality $d\nu_\perp >2$ where $d$ is the space dimensionality. If the Harris criterion is violated,
i.e., if $d\nu_\perp < 2$, weak disorder is relevant because it increases under coarse graining. This means,
the phase transition in the disordered system will be qualitatively different from its clean counterpart.
Two broad cases can be distinguished \cite{MMHF00}: In some systems, the disorder strength reaches a
nonzero but finite value in the limit of infinite length scales. The resulting finite-randomness
critical points show conventional critical behavior, but the critical exponent values differ from the
corresponding clean ones. In contrast, if the disorder strength increases without limit under coarse
graining, the phase transition is controlled by an unconventional infinite-randomness critical point.

The second classification arises from analyzing the physics of rare strong disorder fluctuations and
the spatial regions that support them. Such regions can be locally in one phase while the bulk of the system
is in the other phase. Their contributions to thermodynamic quantities lead to nonanalyticities, now known
as Griffiths singularities \cite{Griffiths69,McCoy69},
not just at the critical point but in an entire parameter region around it.
The character of the Griffiths singularities depends on the effective dimensionality $d_{RR}$ of the rare
regions and on the lower critical dimension $d_c^-$ of the transition at hand. This leads to the following
classification \cite{Vojta06,VojtaSchmalian05}: If $d_{RR} < d_c^-$ (class A), individual rare regions cannot
undergo the phase transition independently from the bulk system. Their slow fluctuations lead to
weak essential Griffiths singularities that are likely unobservable in experiments \cite{Imry77}.
In the opposite case, $d_{RR} > d_c^-$ (class C), individual rare regions can order independently. Long-range
order thus arises gradually rather than via an abrupt collective effect, i.e., the global phase transition
is destroyed by smearing \cite{Vojta03a,Vojta03b}. In the limiting, marginal case $d_{RR}=d_c^-$ (class B), rare regions
cannot yet order, but their dynamics is ultraslow. This leads to enhanced Griffiths singularities,
sometimes dubbed quantum Griffiths or Griffiths-McCoy singularities, that are characterized by power
laws with a nonuniversal Griffiths dynamical exponent $z'$ \cite{ThillHuse95,Fisher95,YoungRieger96}.

These two classifications have been employed to organize the properties of a host of classical, quantum,
and nonequilibrium phase transitions. However, as they focus on different aspects of the randomness,
their predictions sometimes appear to contradict each other. An especially interesting situation occurs
when Harris inequality $d\nu_\perp >2$ is fulfilled, predicting that the disorder is irrelevant while the
rare region classification predicts strong power-law Griffiths singularities. To gain an understanding
of the interplay between the two classifications, it is desirable to investigate a specific example of
such a phase transition.

In this paper, we therefore analyze the nonequilibrium transition of the disordered contact process \cite{HarrisTE74}
in five space dimensions. We will show that its critical behavior is identical to that of the
clean five-dimensional contact process (at least for sufficiently weak disorder), in agreement with the Harris criterion. The critical point
is accompanied by off-critical power-law Griffiths singularities, as predicted by the rare
region classification. However, the Griffiths dynamical exponent $z'$ saturates at a finite value
as the transition is approached, in contrast to the infinite-randomness scenario realized
in the disordered contact process in one \cite{HooyberghsIgloiVanderzande03,VojtaDickison05},
two \cite{OliveiraFerreira08,VojtaFarquharMast09}, and three \cite{Vojta12} space dimensions where $z'$ diverges.
Our paper is organized as follows. We introduce the disordered contact process and discuss its basic properties
in Sec. \ref{sec:CP}. In Sec.\ \ref{sec:OFT}, we develop an optimal fluctuation theory for
the Griffiths singularities. It is based on a recently established
general relation between the Harris criterion and rare region properties \cite{VojtaHoyos14}.
Section \ref{sec:MC} is devoted to large-scale Monte-Carlo simulations
of the clean and disordered five-dimensional contact process on systems with up to
$70^5$ sites that confirm and illustrate our theory.
We conclude in Sec.\ \ref{sec:Conclusions}.

%%%%%%%%%%%%%%%%%%%%%%%%%%%%%%%%%%%%%%%%%%%%%%%%%%%%%%%%%%%%%%%%%%%%%%%%%%%%%%%%%
\section{Contact Process}
\label{sec:CP}
%%%%%%%%%%%%%%%%%%%%%%%%%%%%%%%%%%%%%%%%%%%%%%%%%%%%%%%%%%%%%%%%%%%%%%%%%%%%%%%%%
\subsection{Definition}
%%%%%%%%%%%%%%%%%%%%%%%%%%%%%%%%%%%%%%%%%%%%%%%%%%%%%%%%%%%%%%%%%%%%%%%%%%%%%%%%%

The contact process \cite{HarrisTE74} is a prototypical nonequilibrium many-particle system
that can be understood as a model for the spreading of an epidemic in space.
It can be defined as follows. Each site of a $d$-dimensional hypercubic regular lattice
of $L^d$ sites can be in one of two states, either active (infected) or inactive (healthy).
Over time, active sites can spread the epidemic by infecting their neighbors, or they
can heal spontaneously. To be more precise, the time evolution of the contact process
is a continuous-time Markov process. Infected sites heal spontaneously
at a rate $\mu$ while healthy sites become infected by their neighbors at a rate
$\lambda n /(2d)$. Here, $n$ is the number of sick nearest neighbors of the given site.
The infection rate $\lambda$ and the healing rate $\mu$ are the control
parameters that govern the behavior of the system; without loss of generality
we can set $\mu=1$ fixing the overall time scale.

The basic properties of the contact process are easily understood. If healing dominates over
infection, $\lambda \ll \mu$, the epidemic eventually dies out completely, i.e.,
the system ends up in a state without any active (infected) sites. This state is a
fluctuationless absorbing state that the system cannot leave, it represents the inactive
phase. In contrast, for $\lambda \gg \mu$, the infection will survive to infinite time
(with probability one). In this case, the density of infected sites approaches a nonzero
constant in the long-time limit. This steady state represents the active phase.
The nonequilibrium phase transition between the active and inactive phases, which occurs
at a critical value of the ratio $\lambda/\mu$, belongs to the directed percolation universality
class \cite{GrassbergerdelaTorre79,Janssen81,Grassberger82}.
The order parameter of this transition is the long-time limit of the density of infected sites,
\begin{equation}
\rho_{\rm stat} = \lim_{t\to\infty} \rho(t) = \lim_{t\to\infty} \frac 1 {L^d} \sum_{\mathbf{r}} \langle  n_\mathbf{r}(t) \rangle~.
\label{eq:rho_definition}
\end{equation}
Here, $n_\mathbf{r}(t)$ is the occupation of site $\mathbf{r}$ at time $t$, i.e.,
$n_\mathbf{r}(t)=1$ if the site is infected and $n_\mathbf{r}(t)=0$
if it is healthy.  $\langle \ldots \rangle$ denotes the
average over all
realizations of the Markov process.

%%%%%%%%%%%%%%%%%%%%%%%%%%%%%%%%%%%%%%%%%%%%%%%%%%%%%%%%%%%%%%%%%%%%%%%%%%%%%%%%%
\subsection{Mean-field theory}
\label{subsec:MF}
%%%%%%%%%%%%%%%%%%%%%%%%%%%%%%%%%%%%%%%%%%%%%%%%%%%%%%%%%%%%%%%%%%%%%%%%%%%%%%%%%

The mean-field theory of the clean contact process can be derived by starting from the
Master equation of the contact process and replacing the individual occupation numbers
$n_\mathbf{r}(t)$ by their average $\rho(t)$ (see, e.g., Refs.\ \cite{MarroDickman99,Luebeck04}).
This leads to the differential equation
\begin{equation}
\frac d {d t} \rho(t) = (\lambda-1) \rho(t) -\lambda \rho^2(t)~.
\label{eq:CP_MF}
\end{equation}
For $\lambda<\lambda_c^{MF}=1$, this equation has only one stationary solution, viz.\,
the absorbing state solution $\rho=0$. For $\lambda>1$, there is also
the stationary solution $\rho=(\lambda-1)/\lambda$ representing the active phase.
Thus, $\lambda_c^{MF}$ is the critical point of the nonequilibrium transition within
mean-field approximation.
Close to $\lambda_c^{MF}$, the stationary density varies as $\rho \sim (\lambda-\lambda_c^{MF})^{\beta}$
with $\lambda$. The order parameter exponent takes the mean-field value $\beta=1$.
For $\lambda < \lambda_c^{MF}$, the density decays exponentially with time,
$\rho(t) \sim \exp(-|1-\lambda|t)$. This defines the correlation time $\xi_t = |1-\lambda|^{-1}$.
Comparing with the general definition $\xi_t \sim |\lambda -\lambda_c|^{-\nu_\parallel}$
of the correlation time exponent gives the mean-field value $\nu_{\parallel} = 1$.

Spatial variations can be included in the mean-field theory by treating the density
as a continuum field $\rho(\mathbf{x},t)$ and adding a diffusion term $D\nabla^2 \rho$ to the
mean-field equation (\ref{eq:CP_MF}). Simple dimensional analysis, i.e., comparing the terms
containing the space and time derivatives, gives
a dynamical exponent $z=2$. The correlation length exponent $\nu_\perp$ defined via
the divergence of the spatial correlation length  $\xi \sim |\lambda -\lambda_c|^{-\nu_\perp}$
therefore takes the mean-field value $\nu_\perp=1/2$. For later reference, a summary of
the mean-field critical exponents is given in Table \ref{tab:MF}
(including the initial slip exponent $\Theta$ which will be defined in Sec.\ \ref{sec:MC}).
\begin{table}
\begin{tabular*}{\columnwidth}{l@{\extracolsep\fill}cccccc}
\hline\hline
Exponent         & $\beta$   &   $\nu_\perp$   &  $\nu_\parallel$    &   $z$    &  $\delta$   &   $\Theta$\\
Mean-field value &   1      &       1/2       &        1            &    2     &     1       &       0\\
\hline\hline
\end{tabular*}
\caption{Mean-field critical exponents in the directed percolation universality class. See text for their definitions;
our notation follows Ref.\ \cite{Hinrichsen00}.}
\label{tab:MF}
\end{table}
The mean-field exponents apply if the space dimensionality is above the upper critical dimension $d_c^+ =4$.
If $d<d_c^+$, the exponents are dimension dependent and differ from their mean-field values.

%%%%%%%%%%%%%%%%%%%%%%%%%%%%%%%%%%%%%%%%%%%%%%%%%%%%%%%%%%%%%%%%%%%%%%%%%%%%%%%%%
\subsection{Quenched spatial disorder}
%%%%%%%%%%%%%%%%%%%%%%%%%%%%%%%%%%%%%%%%%%%%%%%%%%%%%%%%%%%%%%%%%%%%%%%%%%%%%%%%%

So far, we have considered the clean contact process which is defined on a regular
periodic lattice and employs uniform rates $\lambda$ and $\mu$ that do not depend on the
lattice site. Quenched spatial disorder can be introduced in a variety of ways.
For example, one can randomly dilute the underlying lattice, or one can make the
infection rate $\lambda_i$ and the healing rate $\mu_i$ independent random functions of the lattice site $i$.
In the following, we will set all healing rates $\mu_i$ to unity as before.
For the infection rates, we will mostly use the binary distribution
\begin{equation}
W(\lambda_i) = p\, \delta(\lambda_i - \lambda_h) + (1-p)\, \delta(\lambda_i - \lambda_l)
\label{eq:binarydistrib}
\end{equation}
with $\lambda_h>\lambda_l$. Here, $p$ is the concentration of large infection rates.

The correlation length exponent $\nu_\perp$ of the contact process takes the values
 1.097 \cite{Jensen99}, 0.733 \cite{VoigtZiff97}, and 0.583 \cite{Vojta12}, in one, two, and three space dimensions, respectively. All
these values violate the corresponding Harris inequality $d\nu_\perp > 2$. Thus, disorder is
a relevant perturbation, and the clean critical behavior is unstable. Detailed studies
of the disordered contact process in one \cite{HooyberghsIgloiVanderzande03,VojtaDickison05},
two \cite{OliveiraFerreira08,VojtaFarquharMast09}, and three \cite{Vojta12} dimensions
showed that the nonequilibrium transition is governed by an unconventional infinite-randomness
critical point.

In contrast, Harris' inequality is fulfilled for the five-dimensional contact process
as $d\nu_\perp =5/2 > 2$. The Harris criterion thus predicts that weak disorder is irrelevant
implying that the disordered five-dimensional contact process should feature the same critical
behavior as the clean one. However, according to the rare region classification put forward in Ref.\
\cite{VojtaSchmalian05} and applied to nonequilibrium transitions in Ref.\
\cite{Vojta06}, the system belongs to class B because the life time of a single
active rare region increases exponentially with its volume. This implies the same type of
power-law Griffiths singularities as in lower dimensions \cite{Noest86,Noest88}.

The five-dimensional disordered contact process is thus indeed a member of the interesting class
of systems for which disorder is perturbatively irrelevant while the nonperturbative rare region
effects are expected to be strong. We will spend the rest of this paper exploring how these two
predictions can be reconciled.

%%%%%%%%%%%%%%%%%%%%%%%%%%%%%%%%%%%%%%%%%%%%%%%%%%%%%%%%%%%%%%%%%%%%%%%%%%%%%%%%%
\section{Optimal fluctuation theory}
\label{sec:OFT}
%%%%%%%%%%%%%%%%%%%%%%%%%%%%%%%%%%%%%%%%%%%%%%%%%%%%%%%%%%%%%%%%%%%%%%%%%%%%%%%%%

In this section, we develop an optimal fluctuation theory for the rare region effects in the
disordered contact process that will allow us to distinguish the cases $d<d_c^+=4$ and $d>d_c^+=4$.
It is an implementation of the ideas developed in Ref.\ \cite{VojtaHoyos14} for the
problem at hand.

%%%%%%%%%%%%%%%%%%%%%%%%%%%%%%%%%%%%%%%%%%%%%%%%%%%%%%%%%%%%%%%%%%%%%%%%%%%%%%%%%
\subsection{Below the upper critical dimension $d_c^+=4$}
\label{subsec:belowdc+}
%%%%%%%%%%%%%%%%%%%%%%%%%%%%%%%%%%%%%%%%%%%%%%%%%%%%%%%%%%%%%%%%%%%%%%%%%%%%%%%%%

We start by considering a large spatial region of linear size $L_{RR}$ containing
$N \sim L_{RR}^d$ lattice sites. Its effective distance from criticality $r_{RR}$ is determined
by the average of the local infection rates $\lambda_i$ over all
sites in the region, $r_{RR}=(1/N)\sum_i \lambda_i -\lambda_c^0$
where $\lambda_c^0$ is the clean bulk critical infection rate.
If the $\lambda_i$ are governed by the binary distribution (\ref{eq:binarydistrib}),
the probability distribution of the rare region distance from criticality is
a binomial distribution,
\begin{equation}
P(r,L_{RR}) = \sum_{n=0}^N \binom N n p^n (1-p)^{N-n} \,\delta \left[r- r_{RR}(N,n) \right] .
\label{eq:binomial}
\end{equation}
with $r_{RR}(N,n)=\lambda_l +\frac n N (\lambda_h - \lambda_l)-\lambda_c^0$.
For large regions of roughly average composition, this binomial
can be approximated by a Gaussian
\begin{equation}
P_G(r,L_{RR}) \sim \exp\left[ - \frac 1 {2b^2}\, L_{RR}^d \,(r-r_{av})^2 \right] ,
\label{eq:gaussian}
\end{equation}
where $r_{av}=p \lambda_h +(1-p) \lambda_l - \lambda_c^0$ is the average distance from criticality
and $b^2=p(1-p)(\lambda_h-\lambda_l)^2$ measures the strength of the disorder.
We are particularly interested in regions that are locally in the active phase,
$r>0$, while the bulk system is still on the inactive side of the transition,
$r_{av}<0$. These (rare) regions are responsible for the Griffiths singularities
in the contact process.

Let us now determine the contribution of the rare regions to the time evolution of the density
$\rho$ of infected sites. To this end, we need to combine the probability distribution (\ref{eq:binomial})
with an estimate of the life time $\tau(r,L_{RR})$ of a single rare region.
If the rare region is locally in the active phase, $r>0$, it can only decay via a
coherent fluctuation of all sites in the region. The probability of such an atypical event
is exponentially small in the rare region volume \cite{Noest86,Noest88}, resulting
in an exponentially large life time $\tau(r,L_{RR}) = \tau_0 \exp[ a L_{RR}^d ]$ and,
correspondingly, in an exponentially small decay rate
\begin{equation}
\epsilon(r,L_{RR}) = [\tau(r,L_{RR})]^{-1} = \epsilon_0 \exp\left[- a L_{RR}^d \right]
\label{eq:tau}
\end{equation}
where $\tau_0 = \epsilon_0^{-1}$ is a microscopic time scale. The coefficient $a$ vanishes at $r=0$
and increases with increasing $r$, i.e., the deeper the region is in the active phase,
the larger $a$ becomes. The functional form of this dependence can be worked out
using finite-size scaling \cite{Barber_review83}.
Below the upper critical dimension, we can use the conventional form of
finite-size scaling. As the coefficient $a$ has the dimension of an inverse volume,
it must scale as $\xi^{-d}$,
\begin{equation}
a = a' r^{d\nu_\perp}~.
\label{eq:a_below}
\end{equation}
The same result also follows from the insight that the term $ a L_{RR}^d$ in
the exponent of (\ref{eq:tau}) represents the number $(L_{RR}/\xi)^d$ of independent correlation volumes
that need to decay coherently. Note that $\nu_\perp$ is the
\emph{clean} correlation length exponent unless the rare region is very close to
criticality (inside the narrow asymptotic critical region).

Consider a system that is overall in the inactive phase, $r_{av} < 0$. We can derive a
rare region density of states, i.e., a probability distribution of the decay rates $\epsilon$, by
summing over all regions that are locally in the active phase, i.e., all regions with
$r>0$. This yields
\begin{equation}
\tilde \rho (\epsilon) \sim \int_0^{\infty} dr \int_0^\infty dL_{RR} \, P(r,L_{RR})\, \delta\left[ \epsilon -\epsilon(r,L_{RR}) \right] .
\label{eq:DOS_integral}
\end{equation}
This integral can be easily evaluated if we use the Gaussian approximation (\ref{eq:gaussian})
of the probability distribution $P$. We first consider the integral over the rare region size
$L_{RR}$. The $\delta$ function fixes the relevant rare region size at
\begin{equation}
L_{RR}^d = (1/a') \, r^{-d\nu_\perp} \ln(\epsilon_0/\epsilon)
\label{eq:L_vs_r}
\end{equation}
Performing the $L_{RR}$ integral, we obtain, up to subleading logarithmic corrections,
\begin{equation}
\tilde\rho(\epsilon) \sim \frac 1 \epsilon \int_0^{\infty} dr \exp\left[ - \frac {1}{2b^2 a'} \, \frac{(r-r_{av})^2}{r^{d\nu_\perp}} \,
            \ln\left(\frac {\epsilon_0} {\epsilon} \right)\right]~.
\label{eq:DOS_integral_r}
\end{equation}
In the limit $\epsilon\to 0$, this integral can be evaluated in saddle point approximation. The saddle-point
equation reads
\begin{equation}
\frac {\partial}{\partial r} \left [ (r-r_{av})^2 \, r^{-d\nu_\perp} \right] =0
\label{eq:saddlepoint_r}
\end{equation}
and gives the solution
\begin{equation}
r_{sp} =  r_{av} d\nu_\perp /(d\nu_\perp -2) ~,
\label{eq:saddlepointvalue_r}
\end{equation}

It is clear that the validity of the saddle-point method depends on the sign of $(d\nu_\perp -2)$. Two cases need
to be distinguished.

(i) In the case $d\nu_\perp <2$, the saddle-point value $r_{sp}$ of the rare region distance from criticality is positive
and thus within the limits of the $r$-integration in eq.\ (\ref{eq:DOS_integral_r}). Inserting the saddle-point value
back into the integral yields a power-law density of states,
\begin{equation}
\tilde\rho(\epsilon) \sim  \epsilon^{d/z'-1}~.
\label{eq:DOS}
\end{equation}
The nonuniversal Griffiths exponent $d/z'$ depends on the disorder strength $b$ and
on the global distance from criticality $r_{av}$ via
\begin{equation}
d/z' = C \, b^{-2} \, r_{av}^{2-d\nu_\perp}
\label{eq:lambda}
\end{equation}
The prefactor is given by $C= 2 (2-d\nu_\perp)^{d\nu_\perp-2}  d\nu_\perp^{-d\nu_\perp} /a'$.
In the literature on Griffiths singularities, the Griffiths exponent $d/z'$ is often called $\lambda$. We will
not use this notation to avoid confusion with the infection rate.
Equation (\ref{eq:lambda}) implies that $d/z'$ vanishes and the Griffiths dynamical exponent $z'$ diverges as
\begin{equation}
z' = (d/C) b^2 r_{av}^{d\nu_\perp-2}
\label{eq:z'}
\end{equation}
as the bulk transition is approached. Equations (\ref{eq:L_vs_r}) and (\ref{eq:saddlepointvalue_r})
also show that the size of the dominating rare regions increases upon approaching the bulk transition while
their distance from criticality decreases. This means close to the bulk transition, the main contribution
to the density of states $\tilde\rho(\epsilon)$ comes from large rare regions not very deep in the active phase. In this regime,
the Gaussian approximation (\ref{eq:gaussian}) of the probability distribution $P$ is well justified.

The density of states (\ref{eq:DOS}) can now be used to calculate various observable quantities.
For example, the time dependence of the density of active sites $\rho(t)$ is simply the Laplace transform
of $\tilde\rho(\epsilon)$ (up to subleading corrections that stem from the logarithmic relation (\ref{eq:L_vs_r})
between the size and decay rate of a rare region). We thus obtain a power-law time dependence
\begin{equation}
\rho(t) \sim \int_0^\infty d\epsilon \tilde\rho(\epsilon) \exp(-\epsilon t) \sim  t^{-d/z'}~
\label{eq:rho_t_Griffiths}
\end{equation}
governed by the Griffiths exponent $d/z'$.

(ii) In contrast, the saddle point method fails if $d\nu_\perp \ge 2$ because $r_{sp}$
either does not exist ($d\nu_\perp=2$) or is negative and thus outside the integration interval ($d\nu_\perp > 2$).
Analyzing the integrand in eq.\ (\ref{eq:DOS_integral_r}) reveals that the exponent attains its
maximum for $r\to \infty$.
The density of states is thus dominated by contributions from the far tail of the probability
distribution $P(r,L_{RR})$, i.e., by small rare regions deep inside the active phase.
In this regime,
the Gaussian approximation (\ref{eq:gaussian}) of the probability distribution is
not justified. Instead, one needs to analyze the tail of the original binomial
distribution.

The far tail of the binomial distribution (\ref{eq:binomial}) consists of regions in
which all sites have the higher of the two infection rates.  For such regions, the binomial
(\ref{eq:binomial})  simplifies to
\begin{equation}
P(r, L_{RR}) \sim \exp(-\tilde p L_{RR}^d) \, \delta (r-\lambda_h+\lambda_c^0)
\label{eq:P_compact}
\end{equation}
with $\tilde p = -\ln(p)$. Combining this with the decay rate $\epsilon(r,L_{RR})$
from eq.\ (\ref{eq:tau}), we find the same power-law density of states
$\tilde \rho(\epsilon) \sim \epsilon^{d/z'-1}$ as in eq.\ (\ref{eq:DOS}), but with the Griffiths
exponent given by
\begin{equation}
 d/z' = \tilde p / a \qquad \textrm{with} \quad a=a'(\lambda_h-\lambda_c^0)^{d\nu_\perp}~.
\label{eq:lambda_compact}
\end{equation}
How does the coefficient $a$ behave close to the phase transition of the disordered system?
At the bulk critical point,
$\lambda_h$ must be larger than $\lambda_c^0$ while $\lambda_l$ must be smaller than $\lambda_c^0$.
This implies that the rare regions, which consist of sites having the high infection rate $\lambda_h$, are (some distance)
inside the active phase. Consequently, $a$ takes the nonzero finite value $a_c=a'(\lambda_h^{cr}-\lambda_c^0)^{d\nu_\perp}$
at the bulk critical point where $\lambda_h^{cr}$ is the value of $\lambda_h$ at the bulk critical point.
The value of $a_c$ depends on how deep the rare region is inside the active phase, it increases with increasing disorder strength.
Consequently, the Griffiths exponent $d/z'$ remains finite at bulk criticality, which implies that
the Griffiths dynamical exponent $z'$ does not diverge; instead it saturates at the
nonuniversal value $z'_c= da_c/\tilde p$ which increases with increasing disorder strength.

Our rare region theory thus establishes a relation between the Harris criterion and the Griffiths
singularities: The same inequality that governs the scaling of the average disorder strength also
controls the Griffiths singularities.  If the clean correlation length exponent
$\nu_\perp$ fulfills the inequality $d\nu_\perp > 2$, the average disorder strength scales to
zero under coarse graining. This means, the clean critical point is stable.
At the same time, the Griffiths dynamical exponent $z'$ takes a
finite value $z'_c$ at the transition point that vanishes for zero disorder and increases with
increasing disorder strength. If $d\nu_\perp <2$, the average disorder strength increases under coarse
graining. This means, the disorder is relevant and destabilizes the clean critical point.
In this case, the Griffiths dynamical exponent $z'$ diverges upon approaching the bulk
critical point.

Let us now discuss the range of validity of this approach. The simple averaging procedure underlying
eqs.\ (\ref{eq:binomial}) and (\ref{eq:gaussian}) corresponds to a tree-level renormalization-group
treatment of the disorder. The theory does not contain any nontrivial disorder renormalizations
beyond tree level. For this reason, the theory correctly describes the behavior
of the disorder close to the clean critical (fixed) point. This means, for $d\nu_\perp>2$, it holds in the
entire critical region. In contrast, it does not describe the asymptotic critical region of a
random fixed point (if any) emerging in the case of $d\nu_\perp <2$.

The limits of our approach for  $d\nu_\perp <2$ can be estimated using scaling arguments.
The scale dimension of the disorder strength $b^2$ at the clean critical point is
$2/\nu_\perp -d$ (see, e.g., Ref.\ \cite{Cardy_book96}). The crossover from the clean
renormalization-group fixed point to the random fixed point is therefore determined
by the value of the scaling combination $b^2 r^{d\nu_\perp -2}$. As long as $b^2 r^{d\nu_\perp -2}$
is small, the behavior is controlled by the clean fixed point. The clean description breaks
down if  $b^2 r^{d\nu_\perp -2}$ reaches a constant of order unity.
According to eq.\ (\ref{eq:z'}), the Griffiths dynamical exponent $z'$ is identical to
this scaling combination (up to a constant prefactor). It thus reaches a value of order unity
before our theory breaks down, independent of the bare disorder strength. The further
evolution of $z'$ in the asymptotic critical region of the random fixed point (if any)
is beyond the scope of our method.

So far, the considerations in this section have been rather general, they should apply to
all disordered nonequilibrium processes for which the rare region decay rate depends
exponentially on their volume (class B of the rare region classification of Refs.\
\cite{VojtaSchmalian05,Vojta06}). Let us now apply the theory to the disordered contact
process. The upper critical dimension of the clean contact process is $d_c^+=4$.
The results of the present section therefore apply to one, two, and three dimensions
for which the clean correlation length exponent $\nu_\perp$ takes the values 1.097 \cite{Jensen99},
0.733 \cite{VoigtZiff97}, and 0.583 \cite{Vojta12}, respectively. All these values
violate the Harris criterion $d\nu_\perp > 2$ which means that the clean critical point
is unstable. According to our results this also suggests that the Griffiths dynamical exponent
$z'$ diverges at the bulk transition. Explicit analytical (strong-disorder renormalization group)
results in one dimension \cite{HooyberghsIgloiVanderzande03,Hoyos08} as well as Monte-Carlo
simulations in one \cite{VojtaDickison05}, two \cite{OliveiraFerreira08,VojtaFarquharMast09},
and three \cite{Vojta12} dimensions agree with these predictions.

%%%%%%%%%%%%%%%%%%%%%%%%%%%%%%%%%%%%%%%%%%%%%%%%%%%%%%%%%%%%%%%%%%%%%%%%%%%%%%%%%
\subsection{Above the upper critical dimension $d_c^+=4$}
\label{subsec:abovedc+}
%%%%%%%%%%%%%%%%%%%%%%%%%%%%%%%%%%%%%%%%%%%%%%%%%%%%%%%%%%%%%%%%%%%%%%%%%%%%%%%%%

The main interest of the present manuscript is the five-dimensional contact process which
is above the upper critical dimension $d_c^+$. We therefore need to investigate how the
optimal fluctuation theory is modified for $d>d_c^+$.

In the derivation of the optimal
fluctuation theory, we have used scaling arguments only once, viz.,
to find the dependence (\ref{eq:a_below})  of the decay coefficient $a$ on the distance from criticality $r$
via finite-size scaling. Above $d_c^+$, conventional finite-size scaling breaks down
because of dangerously irrelevant variables. Instead, many phase transitions feature
a modified version of finite size scaling \cite{Brezin82,KennaLang91}, also dubbed ``$q$-scaling'' \cite{BercheKennaWalter12},
that replaces the usual scaling combination $r L^{1/\nu_\perp}$ (where $L$ is the system size)
with the combination $r L^{q/\nu_\perp}$ where $q=d/d_c^+$
\footnote{Originally, $q$-scaling was thought to apply to periodic boundary conditions
only, and not to the open boundary conditions more relevant for our rare regions. However,
recent work \cite{BercheKennaWalter12} shows that $q$-scaling also holds for open
boundary conditions.}.

The change in finite-size scaling leads to a corresponding change in the relation between
the decay coefficient $a$ and the distance from criticality $r$.
As $a$ has the dimension of an inverse volume, we obtain
\begin{equation}
a = a' r^{d_c^+\nu_\perp}
\label{eq:a_above}
\end{equation}
instead of eq.\ (\ref{eq:a_below}).
Using this relation in the derivation of the optimal fluctuation theory leads to
$r_{sp} = r_{av} d_c^+ /(d_c^+\nu_\perp -2)$. Correspondingly,
$d$ gets replaced by $d_c^+$ in the exponents (\ref{eq:lambda}) and
(\ref{eq:z'}).
The fate of the Griffiths singularities and the
scaling of the average disorder strength are thus governed by different inequalities.
The average disorder strength increases under coarse graining, making the disorder
(perturbatively) relevant if $d\nu_\perp < 2$ while the Griffiths dynamical exponent
$z'$ diverges if $d_c^+ \nu_\perp < 2$.

Let us now apply these general results to the contact process in dimensions $d>4$.
As $\nu_\perp$ takes the mean-field value 1/2, the Harris criterion $d\nu_\perp >2$ is fulfilled, and weak disorder is
perturbatively irrelevant. The finite-size scaling of the directed percolation
transition above $d_c^+=4$ is of $q$-scaling type \cite{LuebeckJanssen05,JanssenLuebeckStenull07}.
As $d_c^+ \nu_\perp = 2$, the Griffiths singularities are dominated by rare regions
in the far tail of the (binomial) probability distribution. The optimal fluctuation theory
thus predicts that the (weakly) disordered contact process in $d>4$ features clean critical
behavior. The accompanying power-law Griffiths singularities are subleading, their
dynamical exponent $z'$ does not diverge but saturates at a finite value $z'_c$ at the bulk
transition point.   $z'_c$ vanishes in the clean limit and increases with increasing disorder
strengths [see discussion after eq.\ (\ref{eq:lambda_compact})].

%%%%%%%%%%%%%%%%%%%%%%%%%%%%%%%%%%%%%%%%%%%%%%%%%%%%%%%%%%%%%%%%%%%%%%%%%%%%%%%%%
\section{Monte-Carlo simulations}
\label{sec:MC}
%%%%%%%%%%%%%%%%%%%%%%%%%%%%%%%%%%%%%%%%%%%%%%%%%%%%%%%%%%%%%%%%%%%%%%%%%%%%%%%%%

%%%%%%%%%%%%%%%%%%%%%%%%%%%%%%%%%%%%%%%%%%%%%%%%%%%%%%%%%%%%%%%%%%%%%%%%%%%%%%%%%
\subsection{Simulation method}
\label{subsec:MC_method}
%%%%%%%%%%%%%%%%%%%%%%%%%%%%%%%%%%%%%%%%%%%%%%%%%%%%%%%%%%%%%%%%%%%%%%%%%%%%%%%%%

We have performed large-scale Monte-Carlo simulations of the clean  and disordered
contact process on a five-dimensional hypercubic lattice
to test the predictions of the optimal fluctuation theory. Our numerical implementation
of the contact process follows Dickman \cite{Dickman99}; it is identical to the one
used in one, two, and three dimensions in Refs.\ \cite{VojtaDickison05,VojtaFarquharMast09,Vojta12}.
The algorithm starts at time $t=0$ from
some configuration of infected  and healthy sites and consists of a sequence of events. During each event
an infected site $i$ is randomly chosen from a list of all $N_a$ infected sites, then a process is selected,
either infection of a neighbor with probability $\lambda_i/(1+ \lambda_i)$ or healing with probability $1/(1+ \lambda_i)$.
For infection, one of the ten neighbor sites is chosen at random. The infection succeeds if this neighbor
is healthy. The time is then incremented by $1/N_a$.
Using this algorithm, we have simulated large systems with sizes of up to $70^5 \approx 1.7\times 10^9$ sites
using periodic boundary conditions.
All results have been averaged over a large number of disorder configurations, precise numbers will be given
below.

We have carried out two different types of simulations. (i) Spreading runs
start from a single active site in an otherwise inactive
lattice; we monitor the survival probability $P_s(t)$, the number of sites $N_s(t)$
of the active cluster, and its (mean-square) radius $R(t)$. At criticality, these quantities are expected
to follow power laws in time, $P_s \sim t^{-\delta}, N_s \sim t^\Theta$, and $R \sim t^{1/z}$.
(ii) We have also performed density decay runs that start from a completely active lattice during which we observe the time evolution
of the density of active sites $\rho(t)$. At criticality, $\rho$ is expected to decay following the same power law,
$\rho \sim t^{-\delta}$, as the survival probability.

%%%%%%%%%%%%%%%%%%%%%%%%%%%%%%%%%%%%%%%%%%%%%%%%%%%%%%%%%%%%%%%%%%%%%%%%%%%%%%%%%
\subsection{Clean five-dimensional contact process}
\label{subsec:MC_clean}
%%%%%%%%%%%%%%%%%%%%%%%%%%%%%%%%%%%%%%%%%%%%%%%%%%%%%%%%%%%%%%%%%%%%%%%%%%%%%%%%%

We have first performed a number of simulation runs of the clean five-dimensional contact process. The purpose
of these calculations is threefold. First, we intend to test our implementation of the contact process.
Second, we wish to confirm the expected mean-field critical behavior. Third, we want to investigate
how the decay rate of a small system depends on its size and the distance from (bulk) criticality. In other words,
we wish to test the predictions of eqs.\ (\ref{eq:tau}) and (\ref{eq:a_above}).

\begin{figure}
\includegraphics[width=8.3cm]{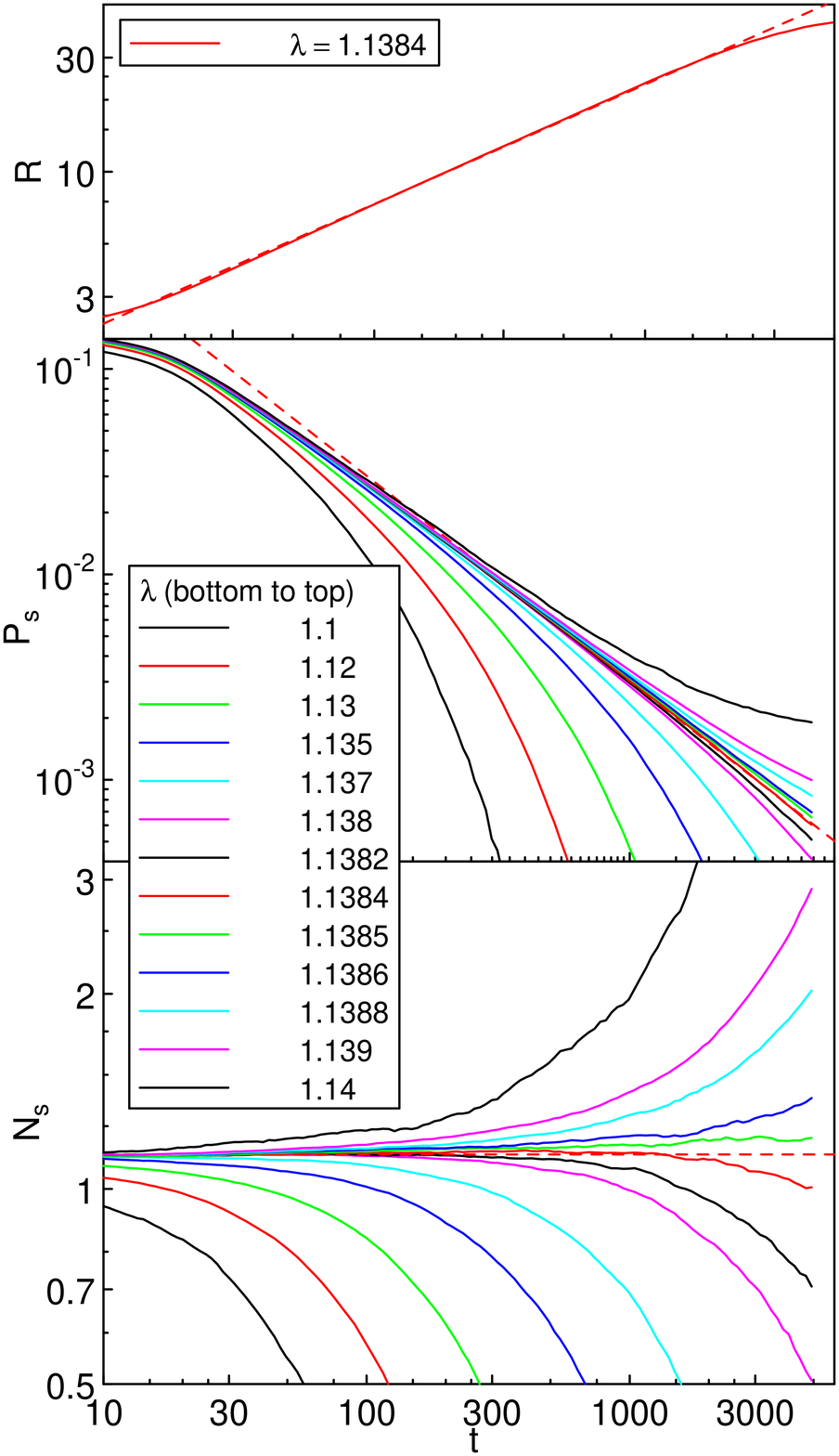}
\caption{(Color online) Spreading simulations for the clean five-dimensional contact process:
Radius $R$ of the active cloud, survival probability $P_s$ and
number of active site $N_s$ vs.\ time $t$ for several infection rates $\lambda$ close to criticality.
The system size is $70^5$ sites; the data are averages over
$2 \times 10^5$ to $6 \times 10^6$ attempts, depending on $\lambda$.
The dashed lines are fits of the data for $\lambda=1.1384$ to mean-field critical behavior, $R \sim t^{1/2}$,
$P_s \sim t^{-1}$, and $N_s \sim t^0$. }
\label{fig:clean_spread}
\end{figure}

Figure \ref{fig:clean_spread} shows the results of spreading simulations (starting from a single active seed site)
on systems of $70^5$ lattice sites.
From these data, we determine the clean critical infection rate to be $\lambda_c^0=1.13845(5)$ where the number in
brackets is an estimate of the error of last digit. The survival probability $P_s$, the number of active sites $N_s$,
and the radius of the active cloud $R$ at this infection rate
can be fitted to the mean-field behavior $P_s \sim t^{-1}$, $N_s \sim t^0$,
and $R\sim t^{1/2}$ discussed in Sec.\ \ref{subsec:MF} with high precision. (In fact, unrestricted power-law fits
give the exponents $\delta=0.99(2)$, $\Theta = -0.005(10)$, and $1/z=0.503(6)$, respectively.) The downward turn of $R(t)$ at the latest times
is due to the fact that the diameter of the active cloud reaches the system size, limiting further growth.
We have therefore restricted our fits to times before that downturn. In addition to the spreading simulations
we have also performed density decay simulations on lattice with $50^5$ sites. They confirm the value of the
critical infection rate as well as the mean-field behavior $\rho \sim t^{-1}$ of the density at criticality.

\begin{figure}
\includegraphics[width=8.3cm]{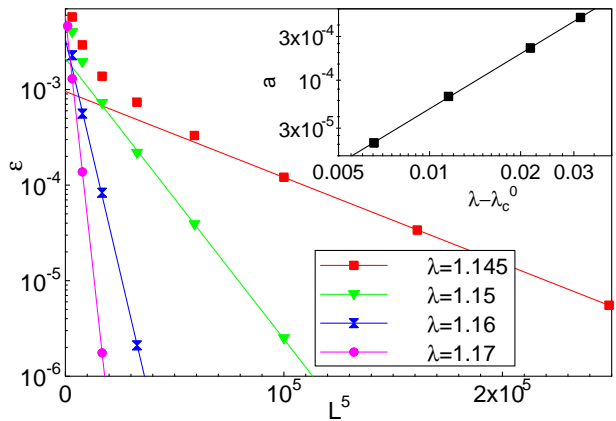}
\caption{(Color online) Semilog. plot of the
decay rate $\epsilon$ vs. system volume $L^5$ for several
infection rates $\lambda$ slightly above the bulk critical rate $\lambda_c^0=1.13845$.
The solid lines are fits to $\epsilon =  \epsilon_0 \exp(- a L_{RR}^5)$.
Inset: Decay coefficient $a$ vs. distance from
bulk criticality $\lambda-\lambda_c^0$. The solid line is a power-law fit
giving the exponent 1.99(3) and the prefactor $a'=0.48(6)$.}
\label{fig:lifetimes}
\end{figure}

To test the predictions  (\ref{eq:tau}) and (\ref{eq:a_above}) for the decay rate $\epsilon$ (and life time $\tau$)
of small systems on the active side of the nonequilibrium transition, we have performed density decay runs
for systems with sizes between $4^5$ and $12^5$ sites for several $\lambda$ slightly above $\lambda_c^0$.
Fits of the density $\rho(t)$ to the expected exponential long-time decay $\rho \sim \exp(-\epsilon t)$
yield the decay rates $\epsilon$. Figure \ref{fig:lifetimes} shows how $\epsilon$ depends on the system
size.
After initial transients for very small systems
\footnote{These transients are due to the crossover from the critical regime
for small $L$ to the active phase for larger $L$.},
the data follow the exponential dependence $\epsilon =  \epsilon_0 \exp(- a L_{RR}^5)$
predicted in eq.\ (\ref{eq:tau}).  The decay coefficient $a$ increases as $(\lambda-\lambda_c^0)^2$
with increasing distance from criticality, as predicted in  eq.\ (\ref{eq:a_above}).

Note that we have used periodic boundary conditions in the simulations leading to
Fig.\ \ref{fig:lifetimes}. In contrast, real rare regions embedded in a nearly critical
bulk have complicated fluctuating boundary conditions that cannot be simulated easily without
simulating the bulk system itself. However, the exponential dependence (\ref{eq:tau}) of
the life time on the rare region volume is a bulk effect and thus independent of the boundary conditions.
Moreover the functional form of the finite-size scaling relation (\ref{eq:a_above}) does not change
when changing the boundary conditions, only the prefactor $a'$ does.
The results in Fig.\ \ref{fig:lifetimes} thus confirm the predicted behavior but the
value of $a'$ resulting from the fit in the inset cannot be expected to be accurate;
instead it provides an upper bound.

%%%%%%%%%%%%%%%%%%%%%%%%%%%%%%%%%%%%%%%%%%%%%%%%%%%%%%%%%%%%%%%%%%%%%%%%%%%%%%%%%
\subsection{Disordered five-dimensional contact process}
\label{subsec:MC_disordered}
%%%%%%%%%%%%%%%%%%%%%%%%%%%%%%%%%%%%%%%%%%%%%%%%%%%%%%%%%%%%%%%%%%%%%%%%%%%%%%%%%

We introduce quenched spatial disorder by making the infection rates $\lambda_i$
independent random variables drawn from the binary distribution (\ref{eq:binarydistrib}).
We parameterize the higher and lower of the two infection rates as $\lambda_h = \lambda$ and
$ \lambda_l=c \lambda$ where $c\le 1$ is a fixed constant while $\lambda$ remains the tuning
parameter of the transition.
To explore the effects of weak and moderately strong disorder, we first set $c=0.1$ or 0.3
and vary the concentration $p$ of the higher rates from 0.8 to 0.2.

\begin{figure}
\includegraphics[width=8.3cm]{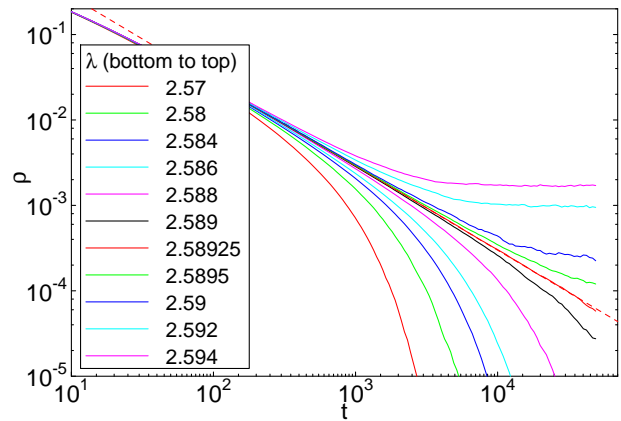}
\caption{(Color online) Density of active sites $\rho$ vs.\ time $t$ for a disordered five-dimensional contact
process with $p=0.2$ and $c=0.3$. The data are averages over 5 to 100 disorder configurations, depending on $\lambda$,
each with $50^5$ lattice sites (one run per disorder configuration). The dashed line represents a  power-law fit of the
critical curve, $\lambda=2.58925(20)$, giving the exponent $\delta=0.99(2)$.}
\label{fig:p02c03}
\end{figure}

Figure \ref{fig:p02c03} shows the results of density decay simulations (starting from a completely
active lattice) for systems of $50^5$ sites with $p=0.2$ and $c=0.3$.
The time dependence of the density of active sites at the critical infection rate of
$\lambda_c=2.58925$ follows the mean-field prediction $\rho \sim t^{-1}$ with high
accuracy.  We have performed analogous density decay simulations for two more
parameter sets, $p=0.5, c=0.3$ and $p=0.8, c=0.1$. In both cases, we find the same
mean-field decay $\rho \sim t^{-1}$ at criticality.

\begin{figure}
\includegraphics[width=8.3cm]{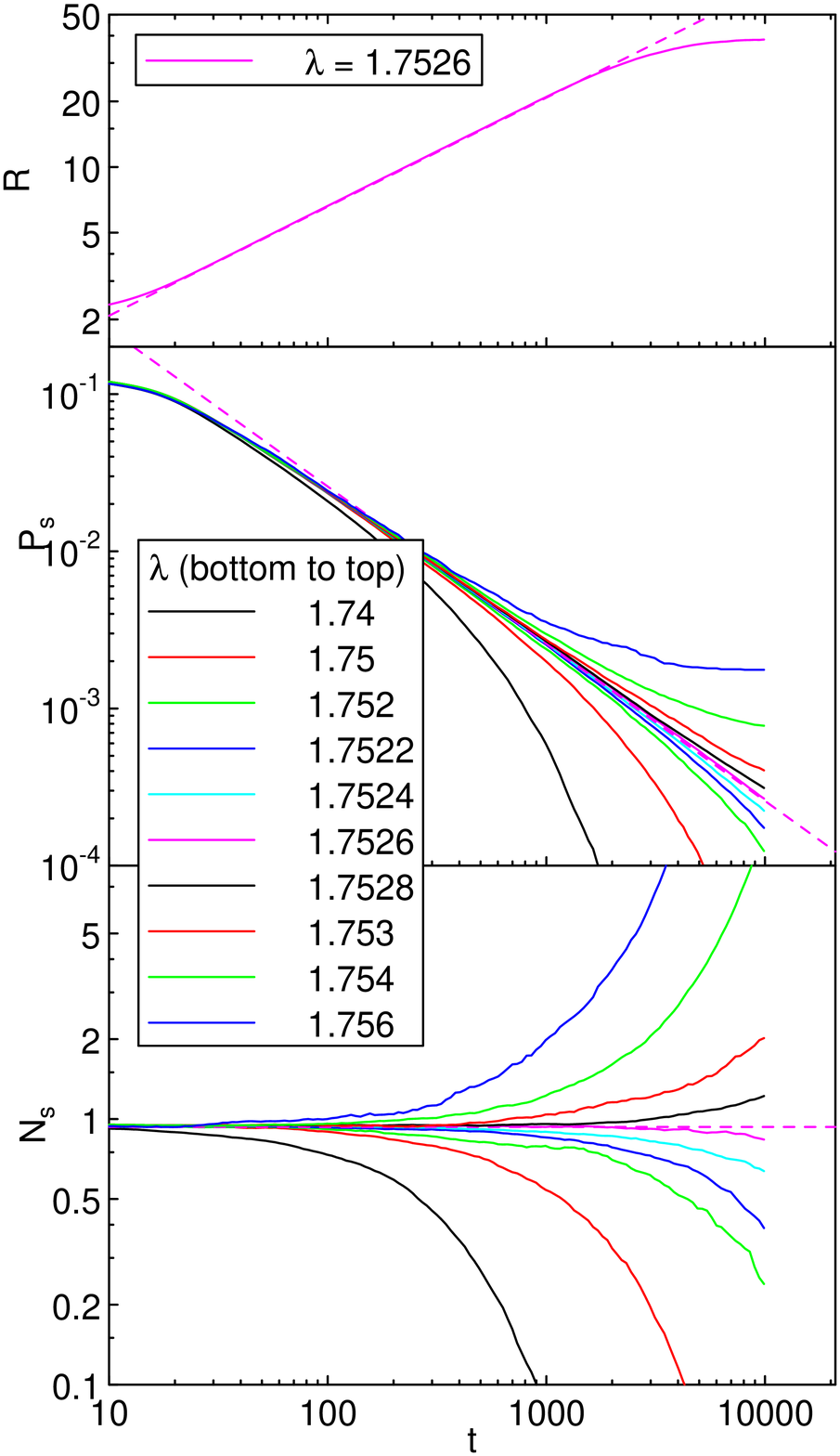}
\caption{(Color online) Number of active sites $N_s$, survival probability $P_s$, and cloud radius $R$  vs.\ time $t$ for a disordered five-dimensional contact
process with $p=0.5$ and $c=0.3$. The data are averages over 1000 to 10000 disorder configurations of $60^5$ sites (1000 trials per disorder configurations). The dashed lines are fits of the data for $\lambda=1.7526$ to mean-field critical behavior, $R \sim t^{1/2}$,
$P_s \sim t^{-1}$, and $N_s \sim t^0$.
}
\label{fig:p05c03}
\end{figure}

In addition to the density decay runs, we have also carried out spreading simulations
on systems of $60^5$ sites for $p=0.5, c=0.3$. There resulting survival probability $P_s$,
number of active sites $N_s$, and cloud radius $R$ are shown in Fig.\ \ref{fig:p05c03}.
At criticality, $\lambda_c=1.7526$, the data follow the mean-field predictions $P_s \sim t^{-1}$ and $N_s \sim t^0$ with
high accuracy.
We thus conclude that the five-dimensional (weakly and moderately) disordered contact process features mean-field critical behavior,
in agreement with the Harris criterion.

What about the power-law Griffiths singularities predicted in Sec.\ \ref{sec:OFT}? The simulations of the
systems discussed so far [disorder parameters ($p=0.2, c=0.3$), ($p=0.5, c=0.3$), and ($p=0.8, c=0.1$)] do not show any trace of
power-law behavior in the Griffiths region, i.e., for infection rates between the clean critical point $\lambda_c^0$
and the critical point $\lambda_c$ of the disordered system. Instead, the survival probability (for spreading simulations)
and the density of active sites (for density decay runs) decay exponentially with time as would be expected in the absence
of Griffiths singularities. We believe the reason why we cannot observe the Griffiths singularities is
that their maximum dynamical exponent $z'_c$ is too small (or, correspondingly, the Griffiths exponent $d/z'$
is too large) in these moderately disordered systems
\footnote{In principle, one can estimate the value of the Griffiths exponent $d/z'$ from eq.\ (\ref{eq:lambda_compact}).
However this requires knowing the coefficient $a'$ for rare regions of arbitrary shape and, importantly, fluctuating
boundary conditions reflecting a nearly critical bulk. As explained at the end of Sec.\ \ref {subsec:MC_clean}, this
value is very hard to obtain.}.
As a result, the Griffiths singularities dominate the
bulk contribution only after very long times which are unreachable within our simulations.

To test this hypothesis, we have studied stronger disorder by setting $p=0.1$ and $c=0.1$. The concentration
$p=0.1$ of strongly infecting sites (having $\lambda_i=\lambda_h=\lambda$) is below the site percolation threshold
$p_c=0.1408$ \cite{Grassberger03}. Establishing long-range order (activity) therefore relies on the weak sites with
infection rates $\lambda_i= c \lambda$. As a result, the critical point $\lambda_c$ is much higher than the
clean value $\lambda_c^0$. This, in turn, puts rare regions consisting of only strong sites ($\lambda_i=\lambda$)
deep in the active phase, increasing their decay coefficient $a$  and with it the Griffiths dynamical
exponent $z'$ [see eq.\ (\ref{eq:lambda_compact})].

\begin{figure}
\includegraphics[width=8.6cm]{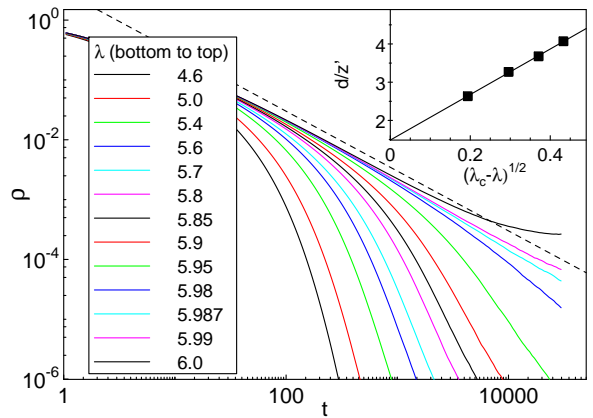}
\caption{(Color online) Density of active sites $\rho$ vs.\ time $t$ for a disordered five-dimensional contact
process with $p=0.1$ and $c=0.1$. The data are averages over 100 to 500 disorder configurations, depending on $\lambda$,
each with $51^5$ lattice sites (one run per disorder configuration). The critical infection rate is
$\lambda \approx 5.987$.
The dashed line represents the mean field power-law with an
arbitrary prefactor. The subcritical curves
($\lambda \le 5.95$) show Griffiths singularities $\rho \sim t^{-d/z'}$
rather than exponential decay. (The curve for $\lambda=5.98$ has not reached the asymptotic regime, yet.)
Inset: Extrapolation of the Griffiths exponent $d/z'$ for $\lambda=5.8 \ldots 5.95$ to criticality (after Ref.\ \cite{VojtaHoyos14}).
 }
\label{fig:p01c01}
\end{figure}

Figure \ref{fig:p01c01} shows results of density decay simulation for systems of $51^5$ sites with
$p=0.1$ and $c=0.1$.
The density decay at the critical infection rate $\lambda_c=5.987$ again follows mean-field behavior $\rho \sim t^{-1}$,
in agreement with the Harris criterion. However, for infection rates slightly below $\lambda_c$, the time dependence
of the density of active sites follows a non-universal power-law, in agreement with eqs.\ (\ref{eq:rho_t_Griffiths})
and (\ref{eq:lambda_compact}). The inset of this figure shows the values of the Griffiths exponent $d/z'$ resulting
from power-law fits of the subcritical $\rho(t)$. Extrapolating $d/z'$ to criticality yields a nonzero finite value,
(in agreement with the prediction of Sec.\ \ref{subsec:abovedc+}.

We have observed analogous subcritical power laws in  simulations of systems with parameters
($p=0.1,c=0.05$) and ($p=0.1,c=0.02$). This raises the interesting question of what happens to the transition if we
further increase the disorder strength by using smaller and smaller values of $c$.
As the strongly infecting sites do not percolate for $p=0.1$, the critical infection rate $\lambda_c$ diverges
in the limit $c \to 0$. Close to criticality, rare regions consisting of only strong sites are thus deeper and
deeper in the active phase, i.e., they have larger and larger decay parameters $a$. Beyond some threshold value of $a$,
the rare region contribution (\ref{eq:rho_t_Griffiths}) to the density will decay \emph{more slowly}
than the bulk mean-field decay $\rho \sim t^{-1}$. It is clear that the critical behavior of such a system must be different
from mean-field behavior.  We emphasize that this results does \emph{not} violate the Harris criterion. The reason is that
the Harris criterion only holds for sufficiently weak disorder as it based on the disorder scaling close to the clean fixed
point. The strong-disorder behavior is beyond the scope of the Harris criterion.
Exploring the novel critical behavior expected for sufficiently strong disorder by numerical means is very
demanding because small values of $c$ lead to extremely slow dynamics. Our simulations of systems with $p=0.1$
and $c$ between 0.001 and 0.05 show indications of non-mean-field behavior. However, within the system sizes accessible
to our simulations ($70^5$ sites), we have not been able to resolve the ultimate fate of these transitions.

%%%%%%%%%%%%%%%%%%%%%%%%%%%%%%%%%%%%%%%%%%%%%%%%%%%%%%%%%%%%%%%%%%%%%%%%%%%%%%%%%
\section{Conclusions}
\label{sec:Conclusions}
%%%%%%%%%%%%%%%%%%%%%%%%%%%%%%%%%%%%%%%%%%%%%%%%%%%%%%%%%%%%%%%%%%%%%%%%%%%%%%%%%

In summary, we have investigated the nonequilibrium phase transition of the five-dimensional contact process
with quenched spatial disorder. This system is a prototypical example of a class of transitions that fulfill
the Harris criterion, predicting clean critical behavior, but also feature strong power-law Griffiths singularities
according to the rare region classification of Refs.\ \cite{Vojta06,VojtaSchmalian05}. To reconcile these predictions,
we have adapted to absorbing state transitions an optimal fluctuation theory recently developed in the context
of quantum phase transitions \cite{VojtaHoyos14}. This optimal fluctuation theory considers the scaling of weak
disorder close to the clean critical point and establishes a relation between the fate of the average disorder
strength and the Griffiths singularities.

For clean critical points below the upper critical dimension $d_c^+$,
both are controlled by the same inequality: If $d\nu_\perp < 2$, weak disorder is relevant and destabilizes the clean
critical behavior while the Griffiths dynamical exponent $z'$ increases with the renormalized
disorder strength upon approaching criticality. In contrast, if $d\nu_\perp >2$, the clean critical behavior is
stable because weak disorder is irrelevant.  At the same time, $z'$ takes a nonzero finite value at the transition.
It is small for weak disorder and increases with the disorder strength.

For clean critical points above the upper critical dimension $d_c^+$, the situation is more complex. Harris' inequality
$d\nu_\perp >2$ still governs the fate of the average disorder strength under coarse graining. However, the behavior of the
Griffiths dynamical exponent $z'$ is controlled by the value of $d_c^+$. If $d_c^+\nu_\perp <2$, the Griffiths dynamical
exponent diverges, if $d_c^+\nu_\perp \ge 2$, it remains finite at the transition point.

The five-dimensional contact process falls into the latter class. Its clean critical point is above the upper
critical dimension $d_c^+=4$. According to the Harris criterion $d\nu_\perp = 5/2 >2$, weak spatial disorder is irrelevant.
Moreover as $d_c^+\nu = 2$, our optimal fluctuation theory predicts the Griffiths dynamical exponent to remain
finite at the transition. For sufficiently weak disorder, the Griffiths singularities thus provide a subleading
correction to the mean-field behavior at criticality. Our Monte Carlo simulations have confirmed these predictions.
We have indeed found mean-field critical behavior over a wide range of disorder strength. For the weakest disorder,
we have not observed any Griffiths singularities. We attribute this to the fact that the Griffiths dynamical exponent
remains very small in these cases, making the Griffiths singularities unobservable within accessible system sizes
and simulation times. We have observed power-law Griffiths singularities for larger disorder. In agreement with the
theoretical predictions, $z'$ extrapolates to a finite value at criticality. For even larger disorder, our simulations
show indications of a change in critical behavior because the Griffiths
singularities become stronger than the mean-field critical singularities.
We note that a similar coexistence of mean-field behavior and Griffiths singularities has also been observed in
the contact process on networks \cite{JOCM12}.

It is instructive to relate the fate of $z'$ at the bulk transition to the geometry of the rare regions.
If $d\nu_\perp \ge 2$ (or, above the upper critical dimension, $d_c^+\nu_\perp \ge 2$), the most relevant
rare regions are small compact clusters deep in the ordered phase. They effectively decouple from the bulk
which explains why the Griffiths dynamical exponent $z'$ is independent of the bulk exponent $z$.
In contrast, for $d\nu_\perp < 2$ (or $d_c^+\nu_\perp < 2$ above the upper critical dimension), the relevant
rare regions become larger and larger as the bulk transition is approached. At criticality they effectively
become indistinguishable from the bulk. As the bulk $z$ is infinite, this explains why $z'$ diverges at
criticality.

Let us conclude by putting our results into the broader perspective of the rare region
classification developed in in Refs.\ \cite{VojtaSchmalian05,Vojta06}. The optimal fluctuation theory developed
in Ref.\ \cite{VojtaHoyos14} and generalized to absorbing state transitions in the present paper applies to class B of
this classification. This class contains systems whose rare regions
are right at the lower critical dimension, $d_{RR}=d_c^-$, leading to power-law Griffiths singularities.
The results of the optimal fluctuation theory allow us to further subdivide class B.
If $d\nu_\perp>2$ (below the upper critical dimension) or if both $d\nu_\perp>2$ and $d_c^+\nu_\perp\ge 2$
(above the upper critical dimension), the system is in class B1 in which clean critical behavior coexists
with subleading Griffiths singularities. The five-dimensional contact process belongs to this subclass,
as does the Ashkin-Teller quantum spin chain discussed in Ref.\ \cite{VojtaHoyos14} (for $\epsilon<-1/2$).
In contrast, if at least one of the inequalities is violated, we expect the critical point to be modified
by the disorder (class B2). In most explicit examples in this subclass such as the transverse-field Ising
model \cite{Fisher92,Fisher95}, itinerant quantum magnets \cite{HoyosKotabageVojta07,VojtaKotabageHoyos09},
or the contact process in $d<4$
\cite{HooyberghsIgloiVanderzande03,Hoyos08,VojtaDickison05,OliveiraFerreira08,VojtaFarquharMast09,Vojta12},
the result is an infinite-randomness critical point, but other strong-disorder scenarios cannot be excluded.

A particularly interesting situation arises above the upper critical dimension if $d\nu_\perp >2$ but
$d_c^+\nu_\perp <2$. The Harris criterion is fulfilled, but our theory suggests dominating Griffiths
singularities because $z'$ becomes large. This opens up the exciting possibility that non-perturbative
rare region physics can modify the transition even if the Harris criterion is fulfilled.
Interestingly, recent strong-disorder renormalization group calculations in $d>4$
\cite{KovacsIgloi11} of the random transverse-field Ising model (for which  $d_c^+\nu_\perp = 3/2 <2$)  show infinite-randomness
criticality even for infinite dimensions.

%%%%%%%%%%%%%%%%%%%%%%%%%%%%%%%%%%%%%%%%%%%%%%%%%%%%%%%%%%%%%%%%%%%%%%%%%%%%%%%%%
\section*{Acknowledgements}
%%%%%%%%%%%%%%%%%%%%%%%%%%%%%%%%%%%%%%%%%%%%%%%%%%%%%%%%%%%%%%%%%%%%%%%%%%%%%%%%%

This work was supported by the NSF under Grant Nos.\ DMR-1205803
and PHYS-1066293, by Simons Foundation, by FAPESP under Grant No.\ 2013/09850-7, and by CNPq under Grant
Nos.\ 590093/2011-8 and 305261/2012-6. We acknowledge the hospitality of the
Aspen Center for Physics.

\bibliographystyle{apsrev4-1}
\bibliography{../00Bibtex/rareregions}
\end{document}